\documentclass[a4paper,aps,twocolumn,amsmath,amssymb,showpacs]{revtex4}
\usepackage{graphicx}
\usepackage{graphics}
\usepackage{psfrag}
\newcommand{\bxi}{\mbox{\boldmath ${\xi}$}}
\newcommand{\bsigma}{\mbox{\boldmath ${\sigma}$}}

\newcommand{\odis}{\big<\hspace{-1mm}\big<}
\newcommand{\cdis}{\big>\hspace{-1mm}\big>}

\newcommand{\Odiss}{\Bigg<\hspace{-2mm}\Bigg<}
\newcommand{\Cdiss}{\Bigg>\hspace{-2mm}\Bigg>}
\begin{document}
\title{Optimal Capacity of the Blume-Emery-Griffiths perceptron}

\author{D. Boll\'e}
\email{Desire.Bolle@fys.kuleuven.ac.be}
\author{I. P\'erez Castillo}
\email{Isaac.Perez@fys.kuleuven.ac.be}
 \affiliation{Instituut voor Theoretische Fysica, K. U. Leuven,  
    B-3001 Leuven, Belgium}
\author{G. M. Shim}
\email{gmshim@cnu.ac.kr}
\affiliation{Department of Physics, Chungnam National University, 
     Yuseong, Daejeon 305-764 R. O. Korea}


\begin{abstract}
A Blume-Emery-Griffiths perceptron model is introduced and its optimal
capacity is calculated within the replica-symmetric Gardner approach, as
a function of the pattern activity and the imbedding stability parameter.
The stability of the replica-symmetric approximation is studied via the
analogue of the Almeida-Thouless line. A comparison is made with other
three-state perceptrons.
\end{abstract}


\maketitle

\section{Introduction.}

Recently an optimal Hamiltonian for a multistate network has been put
forward \cite{DK}, \cite{BV} on the basis of information theory by
maximizing the mutual information content of the system.
For a two-state network, this Hamiltonian equals the well-known Hopfield
Hamiltonian extensively studied in the literature \cite{HKP91},
\cite{MRS95}. For a three-state network one finds a  Blume-Emery-Griffiths
(BEG) spin-glass type Hamiltonian \cite{BEG}. As spin-glasses these
models have been studied for some time now. Thermodynamic as well as
dynamic properties are discussed in the literature for disorder in both
the quadratic and biquadratic interaction. Many references can be found
in \cite{ACN00}. As a neural network model its study has been started
only recently \cite{BV},\cite{BSB}. But it turns out already that both
the maximal capacity and the basin of attraction of this network are
enlarged, at least for Hebb rule learning, in comparison with the
standard three-state networks like, e.g., the Q-Ising spin-glass models.

A natural question is then whether these improved retrieval quality
aspects are restricted to the use of the Hebb rule or whether they are
intrinsic properties of the model. In the same context, a further
question is then whether we can extract a perceptron type model with an
optimal performance out of this BEG recurrent network. The perceptron is
by now a well-known and standard model in theoretical studies and
practical applications in connection with learning and generalization
\cite{HKP91}, \cite{MRS95}, \cite{OK96} -  \cite{EV01}.
Consequently, a number of extensions including many-state, graded
response and colored  perceptrons have been formulated in the literature
\cite{NR91}-\cite{BK01}.

The aim of this work is precisely to introduce such a BEG-perceptron
model and, in particular, to study its Gardner optimal capacity. Although
the method for doing that is standard and well-know by now
\cite{G},\cite{GD} its generalization to the problem at hand is highly
non-trivial. Nevertheless we have succeeded in obtaining a closed
expression for the replica symmetric approximation to the Gardner optimal
capacity.

The paper is organized as follows. In section 2 we recall the BEG
Hamiltonian and define the BEG perceptron model. Section 3 presents a
closed analytic formula for the replica-symmetric Gardner capacity of
this model and studies its behaviour as a function of the imbedding
constant and the activity. Comparisons with other three-state perceptrons
 are made. In section 4 the stability of the replica
symetric solution is studied using an extension of the de Almeida-Thouless
 analysis. The analytic form of the two replicon eigenvalues is obtained.
 Stability is found to be broken for smaller values of the activity and 
for very small imbedding stabilities.
Section 5 presents some concluding remarks.
In the appendices further technical explanations are given.

\section{The BEG perceptron.}
Consider a neural network consisting of $N$ neurons which can take values
$\sigma_i, i=1,\ldots, N$ from the discrete set ${\mathcal S}\equiv
\lbrace -1,0,+1 \rbrace $. The $p$ patterns to be stored in this network
are supposed to be a collection of independent and identically
distributed random variables (i.i.d.r.v.), ${\xi}_i^\mu $, $\mu
=1,\ldots,p$ with a probability distribution
\begin{equation}
p(\xi_{i}^{\mu})=\frac{a}{2}\delta(\xi_{i}^{\mu}-1)+
\frac{a}{2}\delta(\xi_{i}^{\mu}+1)+(1-a)\delta(\xi_{i}^{\mu})
\label{distribution}
\end{equation}
with $a$ the activity of the patterns so that
\begin{equation}
  \lim_{N\rightarrow\infty}\frac{1}{N}\sum_{i}(\xi_{i}^{\mu})^2 = a.
\end{equation}
Given the network configuration at time $t$,
${\bsigma}_N\equiv\{\sigma_j(t)\}, j=1,\ldots,N$, the following dynamics
is considered. The configuration $\bsigma_N(0)$ is chosen as input. The
neurons are updated according to the stochastic parallel spin-flip
dynamics defined by the transition probabilities
\begin{equation}
   \Pr \left(\sigma_i(t+1) = s' \in {\mathcal S}| \bsigma_N(t) \right)
        =
        \frac
        {\exp [- \beta \epsilon_i(s'|\bsigma_N(t))]}
        {\sum_{s \in \mathcal{S}} \exp [- \beta \epsilon_i
                                   (s|\bsigma_N(t))]}\,.
\label{eq:trans}
\end{equation}
Here the energy potential $\epsilon_i[s|{\bsigma}_N(t)]$ is defined by
\begin{equation}
          \epsilon_i[s|{\bsigma}_N(t)] =
          -sh_i({\bsigma}_N(t))-s^2\theta_i({\bsigma}_N(t))
              \,,
\label{eq:energy}
\end{equation}
where the following local fields in neuron $i$ carry all the information
\begin{equation}
        \label{eq:h}
      h_{N,i}(t)=\sum_{j \neq i} J_{ij}\sigma_j(t), \quad
      \theta_{N,i}(t)=\sum_{j\neq i}K_{ij}\sigma_{j}^{2}(t)
\end{equation}
with the obvious shorthand notation for the local fields. For synaptic
couplings $J_{ij}$ and $K_{ij}$ of the Hebb-type
\begin{eqnarray}
&&J_{ij}=\frac{1}{a^{2}N}\sum_{\mu=1}^{p}\xi_{i}^{\mu}\xi_{j}^{\mu}\\
&&K_{ij} = \frac{1}{a^2(1-a)^2N} \sum_{\mu=1}^p
         ((\xi^\mu_i)^2 - a) ((\xi^\mu_j)^2 -a)
\end{eqnarray}
the corresponding neural network Hamiltonian
\begin{equation}
H = - \frac{1}{2} \sum_{i \neq j} J_{ij} \sigma_i \sigma_j
    - \frac{1}{2} \sum_{i \neq j} K_{ij} \sigma_i^2 \sigma_j^2 \ ,
\end{equation}
has been discussed recently \cite{BV}. It has been found that the
capacity and basin of attraction has been enlarged in comparison with
other three-state networks.

We would like to understand whether these better retrieval quality is an
intrinsic property of the model. Therefore, we want to answer the
following question: given the set of $p$ patterns specified above, is
there a network (the best possible network of the BEG-type) which has
these patterns as fixed points of the deterministic form of the dynamics
considered above?  At zero temperature the updating rule  of this
dynamics  (\ref{eq:trans})-(\ref{eq:energy}) is equivalent to the gain
function
\begin{eqnarray}
       \sigma_i(t+1)&=&\mbox{sign}(h_{N,i}(t)) \Theta(|h_{N,i}(t)| 
       + \theta_{N,i}(t))
       \nonumber \\
       &\equiv& \mbox{g}(h_{N,i}(t), \theta_{N,i}(t))
\label{eq:gain}
\end{eqnarray}
with $\Theta$ the Heaviside function. Considering the perceptron 
architecture (N
inputs with couplings $J_j$ and $K_j$ and 1 output) we say that a given
pattern, $\xi_i^\mu, i=1, \ldots, N$,  is stored if there exists a 
corresponding output $\xi_0^\mu$
\begin{equation}
\xi ^\mu_0=g(h ^\mu,\theta ^\mu) \label{metastablecondition}
\end{equation}
with
\begin{equation}
h ^\mu=\frac{1}{\sqrt{N}}\sum_{j=1}^N J_j\xi_j^\mu
\quad\quad 
\theta^\mu =\frac{1}{\sqrt{N}}\sum_{j=1}^N K_j (\xi_j^\mu)^2 
 \, ,
\label{cond2}
\end{equation}
and $\{{\bf J},{\bf  K}\} \equiv \{J_j, K_j\}$ denoting the
configurations in the space of interactions. The factor $N^{-1/2}$ is
introduced to have the weights $J_j$ and $K_j$ of order unity.

The aim is then to determine the maximal number of patterns, $p$, that can be
stored in the perceptron, in other words to find the maximal value of the 
loading $\alpha=p/N$  for which couplings satisfying
(\ref{metastablecondition})-(\ref{cond2}) can still be found. Following a
Gardner-type analysis \cite{G} the fundamental quantity that we want to
calculate is then the volume fraction of weight space  given by
\begin{equation}
V=\int d {\bf J} d{\bf K}\rho({\bf J},{\bf
K})\prod_{\mu=1}^p\chi_{\xi^\mu_0 }(h ^\mu,\theta^\mu ;\kappa)
\label{accesiblevolume}
\end{equation}
with the characteristic function
\begin{eqnarray}
 \chi_{\xi^\mu_0 }(h ^\mu,\theta^\mu ;\kappa) 
  &&=\delta_{\xi ^\mu_0,g(h ^\mu,\theta ^\mu)} \nonumber \\
  &&=(\xi^\mu_0)^2 \Theta(|h^\mu|+\theta ^\mu-\kappa)
              \Theta(\xi^\mu_0  h^\mu -\kappa) \nonumber \\
  &&+(1-(\xi^\mu_0 )^2)\Theta(-|h ^\mu|-\theta ^\mu-\kappa) 
\label{character}
\end{eqnarray}
where $\kappa$ is the imbedding stability parameter measuring the size of
the basin of attraction for the $\mu$-th pattern and  $\rho({\bf J},{\bf
K})$ is the following normalization factor assuming  spherical
constraints for the couplings
\begin{equation}
\rho({\bf J},{\bf K})=\frac{\delta({\bf J}\cdot{\bf J}-N)\delta({\bf
K}\cdot{\bf K}-N)}{\int_{-\infty}^\infty  d {\bf J} d{\bf K}\delta({\bf
J}\cdot{\bf J}-N)\delta({\bf K}\cdot{\bf K}-N)} \, .
 \label{normalization}
\end{equation}
In order to perform the average over the disorder in the input patterns
and the corresponding output we employ the replica technique to evaluate
the entropy per site
\begin{equation}
v =\lim_{N\to\infty}\frac{1}{N}\odis\ln V\cdis \label{entropy}
\end{equation}
where  $\odis\cdots\cdis$ denotes an average over the statistics of
inputs $\{\xi_j^\mu\}$ and outputs $\{\xi_0^\mu\}$, recalling
(\ref{distribution}).

\section{Replica symmetric analysis}

In the replica approach the entropy per site $v$ is computed via the
expression
\begin{equation}
v=\lim_{N\to\infty}\lim_{n\to 0}\frac{1}{nN}\big(\odis
V^n\cdis-1\big)=\lim_{N\to\infty}\lim_{n\to 0}\frac{1}{nN}\ln \odis
V^n\cdis
\end{equation}
where $V^n$ is the $n$-times replicated fractional volume
\begin{widetext}
\begin{equation}
\odis V^n\cdis \propto \int \Big[\prod_{\alpha=1}^n d {\bf J}^\alpha d
{\bf K}^\alpha \delta\Big({\bf J}^\alpha\cdot{\bf
J}^\alpha-N\Big)\delta\Big({\bf K}^\alpha\cdot{\bf
K}^\alpha-N\Big)\Big]\Odiss\prod_{\alpha=1}^n\prod_{\mu=1}^p\chi_{\xi^\mu_0
}(h ^\alpha_\mu,\theta ^\alpha_\mu;\kappa)\Cdiss 
\label{fracvol}
\end{equation}
\end{widetext}
whereby we can forget, since the couplings are continuous, about constant
terms such as the  denominator in (\ref{normalization}). The calculation
then proceeds in a standard way although the technical details are much
more complicated. For a short account we refer to  Appendix A. Here we
restrict ourselves to the following important remarks.
The main order parameters appearing in the calculation are
\begin{eqnarray}
&&q_{\alpha\beta}=\frac{1}{N} {\bf J}^\alpha\cdot{\bf J}^\beta,\quad
r_{\alpha\beta}=\frac{1}{N} {\bf K}^\alpha\cdot {\bf K}^\beta, \quad
\alpha < \beta \nonumber\\
&&L^\alpha=\frac{1}{\sqrt{N}}\sum_{j=1}^N K^\alpha_j,
      \quad \forall \alpha.
      \label{orderpar}
\end{eqnarray}
Of course, in the replica symmetric (RS) approximation we are focussing
upon here, $q_{\alpha\beta}=q, r_{\alpha\beta}=r,L^\alpha=L$. The first
two order parameters are the overlaps between two distinct replicas for
the couplings ${\bf J}$ and ${\bf K}$, the third one arises from the fact
that the dynamics (\ref{eq:gain}) and, hence, also the characteristic
function (\ref{character}), contains a second field $\theta$, quadratic
 in the patterns.
We remark that it describes the relative importance of the active versus 
the non-active neurons.
Actually, in the calculation $aL$ will be the important quantity with $a$
the second moment of the pattern distribution, i.e., the pattern activity.

The RS optimal Gardner capacity is obtained when the overlap order
parameters $q$ and $r$ go to $1$. It is clear that these limits have to
be taken simultaneously but, in general, their rate of convergence could
be different. Therefore, we introduce $(1-r)=\gamma(1-q)$ where $\gamma$
is a new parameter which one also needs to extremize. We expect this
parameter $\gamma$ to depend on the pattern distribution through the
activity $a$.

Pursuing this approach then leads to
\begin{equation}
\alpha_{RS}(a,\kappa)=-\underset{L,\gamma}{\text{extr}}\lim_{q\to1}
                 \frac{1+1/\gamma}{2(1-q)g^{RS}_1(q,\gamma,L)}
          \label{capacity_RS2}
\end{equation}
where $g^{RS}_1(q,\gamma,L)$ reads
\begin{widetext}
\begin{equation}
  g^{RS}_1(q,\gamma,L)=\int {\cal D}(h_0)
   {\cal D}\left(\sqrt{\gamma}\theta_0-l\right)
       \Odiss\ln
    \int_{\Omega_\xi}\frac{dh}{\sqrt{2\pi(1-q)}} \frac{d\theta}
   {\sqrt{2\pi(1-q)}}
   \exp\Big[-\frac{(h-h_0)^2+(\theta-\theta_0)^2}{2(1-q)}\Big]
   \Cdiss_{\xi_0}
\label{gRS2}
\end{equation}
\end{widetext}
with $l \equiv aL/\sqrt{a(1-a)}$, where
 ${\cal D}(ax +b) = (2\pi)^{-1/2}adx \exp[(-1/2)(ax+b)^{2}] $ and
where the integration region $\Omega_\xi$ is determined by the Heaviside
functions appearing in the characteristic function
$\chi_\xi(\sqrt{a}h,\sqrt{\gamma a(1-a)}\theta;\kappa)$ defined in
(\ref{character}).
The expression  (\ref{gRS2}) for the function $g^{RS}_1$  suggests that
an asymptotic expansion to compute the limit $q\to 1$ is possible.
Indeed, after some  tedious algebra (see Appendix B) we find for this
limit
\begin{widetext}
\begin{eqnarray}
g^{RS}_1(q,\gamma,L)&&= -\frac{a}{2(1-q)}\sum_{i=1}^3\int_{{\cal R}_{i}}
 {\cal D}(h_0+\kappa/\sqrt{a}{\cal D}(\sqrt{\gamma}\theta_0-l) \,
            d_{min}^{{\cal R}_{i}}(h_0,\theta_0)\nonumber \\
 &&-\frac{(1-a)}{2(1-q)}\sum_{i=1}^3\int_{{\cal R}'_{i}}
   {\cal D}(h_0) {\cal D}(\sqrt{\gamma}\theta_0-u) \,
     d^{{\cal R}'_{i}}(h_0,\theta_0)+  o(1/(1-q))
\label{Divergence}
\end{eqnarray}
\end{widetext}
with $u\equiv (aL+\kappa)/\sqrt{a(1-a)}$.
The integration regions read 
\begin{eqnarray}
{\cal R}_1&&=\left\{\begin{array}{l}
h_0<0\\
\theta_0>0
\end{array}\right.\label{Region1}\\
{\cal R}_2&&=\left\{\begin{array}{l}
h_{0}\gamma'<\theta_{0}<0\\
h_{0}<0
\end{array}\right.\label{Region2}\\
{\cal R}_3&&=\left\{\begin{array}{l}
\theta_{0}<0\\
\theta_{0}/\gamma'<h_{0}<-\theta_{0}\gamma'\\
\end{array}\right.\label{Region3}\\
{\cal R}'_1&&=\left\{\begin{array}{l}
h_{0}>0\\
-h_0/\gamma'<\theta_{0}<\gamma' h_{0}
\end{array}\right.\label{Region1prim}\\
{\cal R}'_2&&=\left\{\begin{array}{l}
-\theta_0/\gamma'<h_{0}<\theta_0/\gamma'\\
\theta_{0}>0
\end{array}\right.\label{Region2prim}\\
{\cal R}'_3&&=\left\{\begin{array}{l}
h_{0}<0\\
h_0/\gamma'<\theta_{0}<-\gamma' h_{0}
\end{array}\right.\label{Region3prim}
\end{eqnarray}
and the corresponding integrands are given by
\begin{eqnarray}
&&d^{{\cal R}_1}_{min}=h_0^2\label{Mindis1}\\
&&d^{{\cal R}_2}_{min}=h_0^2 +\theta_0^2\label{Mindis2}\\
&&d^{{\cal R}_3}_{min}=\frac{1}{1+(\gamma')^2}\big(h_0+ 
        \gamma'\theta_0\big)^2\label{Mindis3}\\
&&d^{{\cal R}'_1}_{min}=\frac{1}{1+(\gamma')^2}\big(h_0+
         \gamma'\theta_{0}\big)^2\label{Mindis1prim}\\
&&d^{{\cal R}'_2}_{min}=h^2_{0}+\theta_{0}^2\label{Mindis2prim}\\
&&d^{{\cal R}'_3}_{min}=\frac{1}{1+(\gamma')^2}\big(h_0-
        \gamma'\theta_{0}\big)^2\label{Mindis3prim}
\end{eqnarray}
with $\gamma'\equiv \sqrt{\gamma(1-a)}$ and where we remark that the
$d_{min}$ are minimal distances between a point in the different
integration regions ${\cal R}_i,{\cal R'}_i, i=1,2,3$  and the border of
 $\Omega_\xi$(see Appendix B).
This may allow for a possible geometrical interpretation of the Gardner
optimal capacity in the space of local fields  as it has been suggested for 
the $Q$-state clock model in \cite{GBK}.

\begin{figure}
\includegraphics[width=0.38\textwidth,height=0.28\textheight]{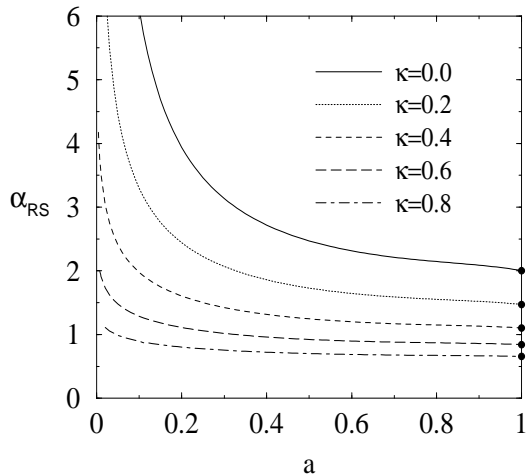}\\
\caption{The optimal capacity $\alpha_{RS}$ as a function of the pattern activity $a$ for several values of the stability constant $\kappa$. The dots at $a=1$ refer to the optimal capacity of the two-state perceptron.}
\label{figure1}
\end{figure}

\begin{figure}
\includegraphics[width=0.38\textwidth,height=0.28\textheight]{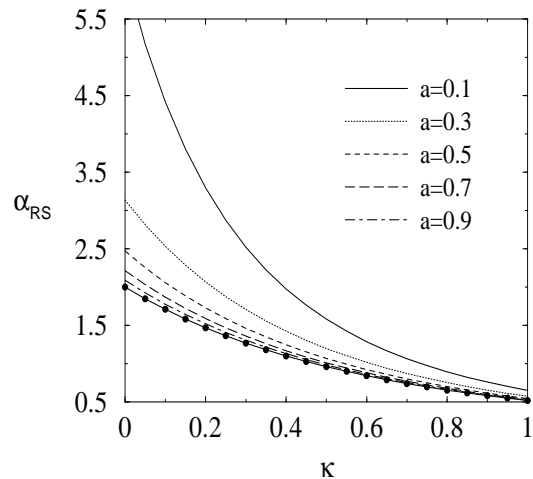}\\
\caption{The optimal capacity $\alpha_{RS}$ as a function of the stability $\kappa$ for several
 values of the pattern activity $a$. The straight-dotted line corresponds to the optimal capacity
  of the two-state perpectron.}
\label{figure2}
\end{figure}

\begin{figure}
\includegraphics[width=0.38\textwidth,height=0.28\textheight]{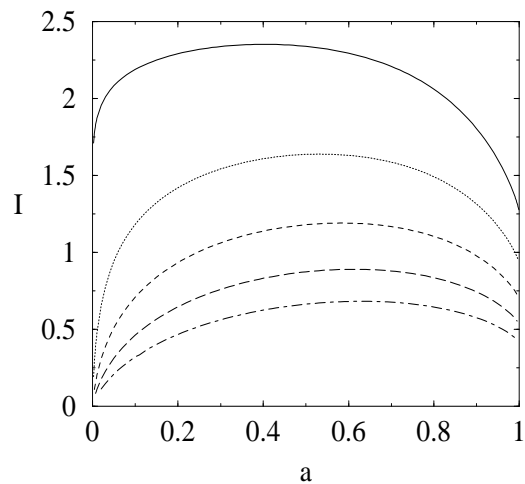}\\
\caption{The information content per neuron, $I$, as a function of $a$ for $\kappa =0;0.2;0.4;0.6;0.8$ (from top to bottom)} 
\label{figure3}
\end{figure}

After inserting \eqref{Divergence}-\eqref{Mindis3prim} in 
\eqref{capacity_RS2} and
extremizing numerically with respect to $L$ and $\gamma$, we find the
results presented in figures \ref{figure1}-\ref{figure2}.
In fig. 1 the capacity $\alpha_{RS}$ versus the activity $a$ is shown for
several values of the imbedding stability constant $\kappa$. For bigger
$\kappa$, the capacity becomes, of course, smaller.  For $a=1$, i.e., binary patterns, we find
back the original Gardner results, as we do in fig. 2 showing
$\alpha_{RS}$ as a function of $\kappa$ for several values of $a$.
Smaller activity indicating a growing presence of zero-state neurons
leads to bigger capacities. Of course, this does not mean a priori that
also  the information content of the system is increased. For
completeness, we 
remark that the parameters $l= aL/\sqrt{a(1-a)}$ and $\gamma$ that we have 
extremized over, depend rather strongly but smoothly on the pattern activity.
For $a=1$ we find back the two-state perceptron value for $L$, i.e. $L=0 
(l=\infty)$, and $\gamma=\infty$.
Finally, in order to have an idea about the information stored into the network we plot in fig. \ref{figure3} the information content per neuron 
\begin{equation}
I(a)= -\frac{\alpha_{RS}}{\ln 3}\left[a \ln(\frac{a}{2}) + (1-a) \ln(1-a)\right].
\end{equation}
For $a=1$ our result is again consistent with the simple perceptron result \cite{G}.
Comparing with other three-state neuron perceptron models we recall that for $\kappa=0$ and uniform patterns 
the $Q=3$ Ising perceptron can maximally reach an optimal capacity equal to $1.5$,  depending on the separation
 between the plateaus of the gain function  (see \cite{MKB91}, \cite{BDM91}) for the precise details) and the 
$Q=3$ clock and Potts model both reach an optimal capacity of $2.40$ \cite{GK94},\cite{GBK} while the value 
for the BEG perceptron found here is $2.24$. Here we have to recall that the $Q=3$ Ising perceptron and the 
BEG perceptron have the same topology structure in the neurons, whereas the $Q=3$ clock and Potts models have 
a different topology.

\section{Stability of the replica symmetric solution}
From the work of Gardner \cite{G} we know that for the binary neuron perceptron
the RS solution is marginally stable against RS breaking (RSB) fluctuations.
From the work on multi-state Q-Ising neurons \cite{BM94} we know that the RS
solution may be stable or unstable depending  on the gain parameter, the number
of spin states and the distribution of the patterns. Furthermore, in
general, increasing the imbedding stability parameter $\kappa$ lowers the
capacity and enhances the stability against RSB. Using these results for
the  $Q=3$
spin states as a guide we also expect breaking for the BEG perceptron model at
hand. To confirm  this and find out the precise interval of $a$ values
where breaking occurs, we generalize the de Almeida-Thouless analysis
\cite{AT78}, \cite{MPV}.

\begin{figure}
\includegraphics[width=0.38\textwidth,height=0.28\textheight]{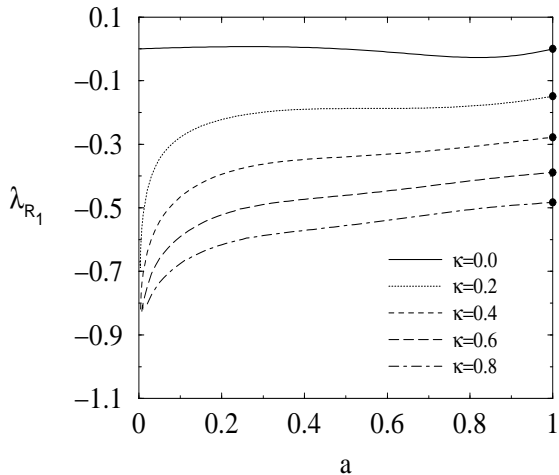}
\caption{The first replicon eigenvalue $\lambda_{R_1}$ as a function of $a$ for several values of $\kappa$. The dots at $a=1$ refer to the optimal capacity of the two-state perceptron.}
\label{figure4}
\end{figure}

\begin{figure}[h]
\includegraphics[width=0.38\textwidth,height=0.28\textheight]{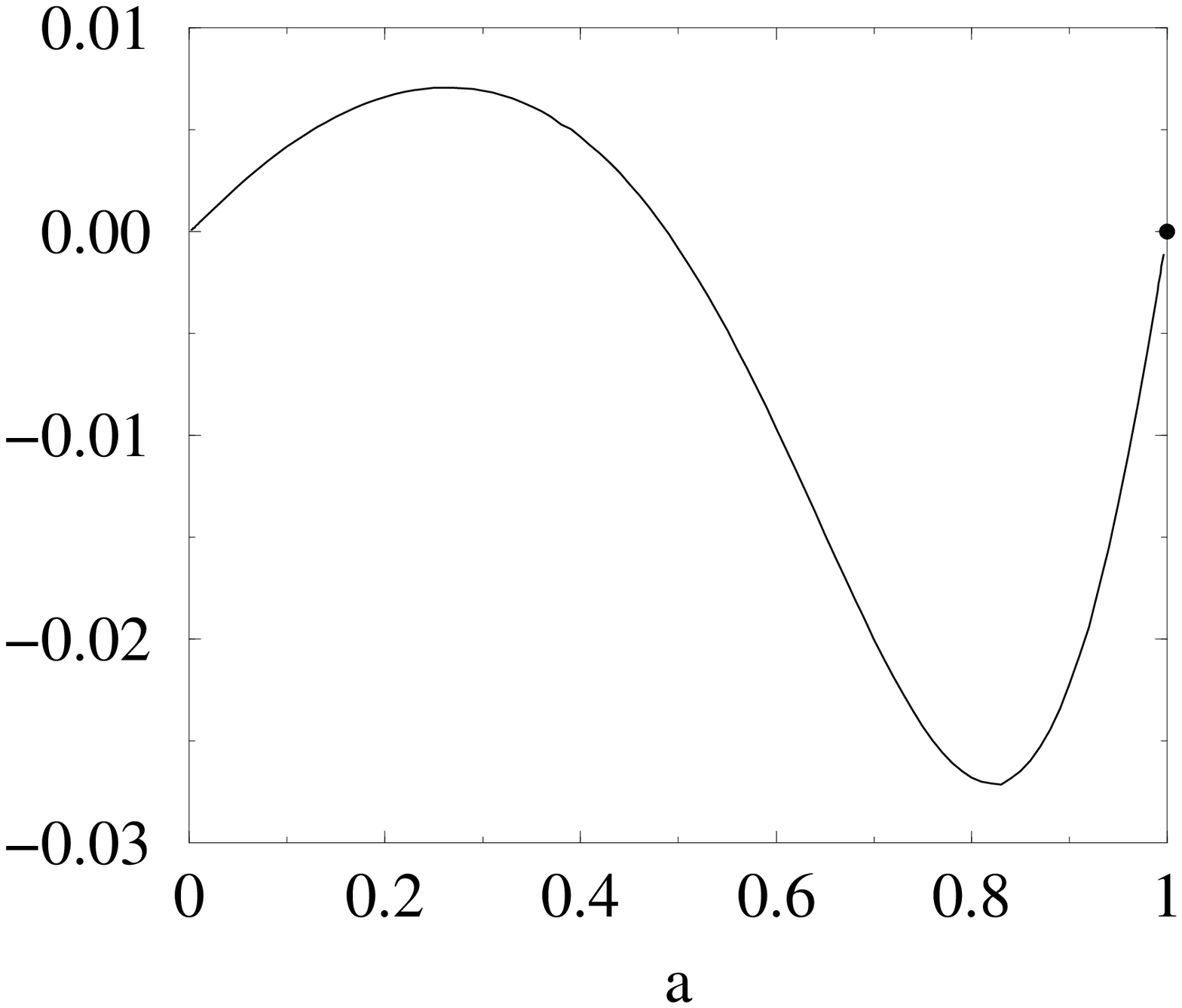}
\caption{The first replicon eigenvalue $\lambda_{R_1}$ as a function of $a$ for $\kappa=0$ on a different scale. RSB occurs for smaller values of $a$.}
\label{figure5}
\end{figure}

\begin{figure}[h]
\includegraphics[width=0.38\textwidth,height=0.28\textheight]{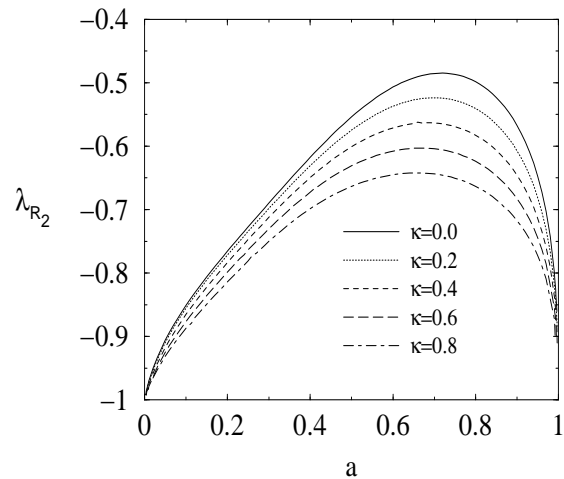}
\caption{The second replicon eigenvalue $\lambda_{R_2}$ as a function of $a$ for several values of $\kappa$.}
\label{figure6}
\end{figure}

First, the hessian matrix associated with the function $\Phi$, eq. 
\eqref{Phifunction}, is computed, and then the eigenvalues are
determined. As usual, two types of eigenvalues are
found: longitudinal eigenvalues describing fluctuations within RS and
transverse eigenvalues describing stability against RSB. We find {\it four}  
transversal eigenvalues each with degeneracy $\frac{1}{2}n(n-3)$. In the limit $q \to 1$
 they can be calculated explicitly in terms of the minimal 
distances occuring in (\ref{Mindis1})-(\ref{Mindis3prim}). The result reads
(for more details we refer to Appendix C) 
\begin{widetext}
\begin{eqnarray}
&& \hspace*{-0.5cm} \lambda_+=\frac{1}{2}(\Delta_{q}
        +\Delta_{r})+\frac{1}{2}\sqrt{(\Delta_{q}-\Delta_{r})^2
        +4\Delta_{c}^2} \label{lplus}\\
&&\hspace*{-0.5cm}\lambda_-=\frac{1}{2}(\Delta_{q}
        +\Delta_{r})-\frac{1}{2}\sqrt{(\Delta_{q}-\Delta_{r})^2
        +4\Delta^2_{c}}\\
&&\hspace*{-0.5cm}\tau_+=\frac{1}{2(\Delta_{c}^2
        -\Delta_{q}\Delta_{r})}\Big\{\Delta_{q}
        +\Delta_{r}+(\Delta_{\widehat{q}}+\Delta_{\widehat{r}})(\Delta_{c}^2
        -\Delta_{q}\Delta_{r})   +\sqrt{4\Delta_{c}^2
        +\big[\Delta_{q}-\Delta_{r}+(\Delta_{\widehat{q}}
        -\Delta_{\widehat{r}})(\Delta_{q}\Delta_{r}-\Delta_{c}^2)\big]^2}\Big\}\\
&&\hspace*{-0.5cm}\tau_-=\frac{1}{2(\Delta_{c}^2
        -\Delta_{q}\Delta_{r})}\Big\{\Delta_{q}
        +\Delta_{r}+(\Delta_{\widehat{q}}+\Delta_{\widehat{r}})(\Delta_{c}^2
        -\Delta_{q}\Delta_{r})-\sqrt{4\Delta_{c}^2
        +\big[\Delta_{q}-\Delta_{r}+(\Delta_{\widehat{q}}
        -\Delta_{\widehat{r}})(\Delta_{q}\Delta_{r}-\Delta_{c}^2)\big]^2}\Big\}
\label{tmin}
\end{eqnarray}
\end{widetext}
with the $\Delta$'s given by
\begin{widetext}
\begin{eqnarray}
\Delta_q&&=\frac{a\alpha_{RS}}{(1-q)^2}\sum_{i=1}^3\int_{{\cal R}_i}{\cal D}(h_0+\kappa/\sqrt{a},\sqrt{\gamma}\theta_0-t)\Big\{\frac{1}{2}\frac{\partial^2}{\partial h_0^2}d^{{\cal R}_i}_{min}(h_0,\theta_0)\Big\}^2
\nonumber \label{delq} \\
&& +\frac{(1-a)\alpha_{RS}}{ (1-q)^2}\sum_{i=1}^3\int_{{\cal R}_i'}{\cal D}(h_0,\sqrt{\gamma}\theta_0-u)\Big\{\frac{1}{2}\frac{\partial^2}{\partial h_0^2}d^{{\cal R}_i'}_{min}(h_0,\theta_0)\Big\}^2+o(1/(1-q))
  \\
\Delta_r&&=\frac{a\alpha_{RS}}{\gamma2 (1-q)^2}\sum_{i=1}^3\int_{{\cal R}_i}{\cal D}(h_0+\kappa/\sqrt{a},\sqrt{\gamma}\theta_0-t)\Big\{\frac{1}{2}\frac{\partial^2}{\partial \theta_0^2}d^{{\cal R}_i}_{min}(h_0,\theta_0)\Big\}^2
\nonumber \\
&&+\frac{(1-a)\alpha_{RS}}{\gamma2 (1-q)^2}\sum_{i=1}^3\int_{{\cal R}_i'}{\cal D}(h_0,\sqrt{\gamma}\theta_0-u)\Big\{\frac{1}{2}\frac{\partial^2}{\partial \theta_0^2}d^{{\cal R}_i'}_{min}(h_0,\theta_0)\Big\}^2+o(1/(1-q))
      \\
\Delta_c&&=\frac{a\alpha_{RS}}{\gamma (1-q)^2}\sum_{i=1}^3\int_{{\cal R}_i}{\cal D}(h_0+\kappa/\sqrt{a},\sqrt{\gamma}\theta_0-t)\Big\{\frac{1}{2}\frac{\partial^2}{\partial h_0\partial \theta_0}d^{{\cal R}_i}_{min}(h_0,\theta_0)\Big\}^2
\nonumber \\
&& +\frac{(1-a)\alpha_{RS}}{\gamma (1-q)^2}\sum_{i=1}^3\int_{{\cal R}_i'}{\cal D}(h_0,\sqrt{\gamma}\theta_0-u)\Big\{\frac{1}{2}\frac{\partial^2}{\partial
h_0\partial \theta_0}d^{{\cal R}_i'}_{min}(h_0,\theta_0)\Big\}^2+o(1/(1-q))\\
\Delta_{\widehat{q}}&&=(1-q)^2\\
\Delta_{\widehat{r}}&&=(1-r)^2=\gamma^2(1-q)^2 \, .
     \label{delr}
\end{eqnarray}
\end{widetext}
Then {\it two} replicon  eigenvalues, $\lambda_{R_1}$ and $\lambda_{R_2}$, can be defined as 
\begin{equation}
\lambda_{R_1}=\lambda_+\tau_-\qquad\lambda_{R_2}=\lambda_-\tau_+
\end{equation}
Stability of the RS solution requires that both
$\lambda_{R_1},\lambda_{R_2}<0$. 
In fig. \ref{figure4}-\ref{figure6} we present the numerical results concerning 
the stability analysis. In fig. \ref{figure4} the first replicon eigenvalue 
$\lambda_{R_1}$ is shown as a function of $a$ for several values of $\kappa$.
 It is seen that for small values of $\kappa$ this eigenvalue becomes positive
  for smaller values of $a$ and hence replica symmetry is broken. We remark
   that for $a=1$ our results are consistent with those of Gardner \cite{G}.
 Fig. \ref{figure5} presents a closer view of this for $\kappa=0$. 
 For $0< a \leq 0.48(8)$ the RS solution is unstable. Storing only zero-state
  spins, $a=0$, or binary spins $a=1$ leads to marginal stability. As a first 
  explanation one could remark that for increasing $a$, allowing more $\pm$
   states, the disorder is increased up to about a uniform distribution of 
   patterns, $a=2/3$. It is clear that for bigger $\kappa$, the stability against RSB increases. In fact  for $\kappa > 0.0061$ already no more breaking occurs. 
Finally, fig. 6 shows that $\lambda_{R_2}$ is always negative and, hence, plays no role in the breaking of the RS stability.

\section{Concluding remarks.}

In this work we have introduced a perceptron model based upon the recently studied Blume-Emery-Griffiths neural network, containing ternary neurons. We have obtained an analytic formula for the replica symmetric optimal Gardner capacity. For the imbedding stability constant equal to zero and uniform patterns, e.g.,  we find a bigger optimal capacity, $\alpha_{RS}=2.24$, than the one for the $Q=3$ Ising perceptron, $\alpha_{RS}=1.5$, which has the same topology structure for the neurons. Since, in general, perceptrons turn out to be very useful models in connection with learning and generalization this is an interesting observation. It is also consistent with earlier results derived for the Hebb rule.

We have studied the stability of the replica-symmetric solution by generalizing the de Almeida-Thouless analysis and deriving an analytic expression for the {\it two} replicon eigenvalues that play a role in the Gardner limit. Breaking only occurs for small activities and very small imbedding constants, $\kappa < 0.0061$. This is consistent with the stability results found for the $Q=3$ Ising perceptrons. 

These results strenghten the idea that the better retrieval properties found for the Blume-Emery-Griffiths model in comparison with the $Q=3$ Ising model are not restricted to the specific Hebb rule but are intrinsic to the model.

\begin{acknowledgments}
We would like to thank Dr. P. Kozlowski and T. Verbeiren for critical discussions. One of us (IPC) especially thanks J. van Mourik and J. Gheerardyn for useful remarks. This work has been supported in part by the Fund of Scientific Research, Flanders-Belgium.
\end{acknowledgments}

\appendix
\section{Replica  analysis and replica symmetric ansatz.}
In this appendix we outline the main steps in the calculation of the n-times replicated volume (\ref{fracvol}) extending  \cite{G} to the case at hand. In order to perform the quenched average we use the $\delta$-function representation 
\begin{widetext}
\begin{eqnarray}
1=\int_{-\infty}^\infty \frac{d h^\alpha_\mu d\widehat{h}^\alpha_\mu}{2\pi}\exp\Big[i\widehat{h}^\alpha_\mu\Big(h^\alpha_\mu-\frac{1}{\sqrt{N}}\sum_{j=1}^N J^\alpha_j \bxi^\mu_j\Big)\Big]\\
1=\int_{-\infty}^\infty \frac{d \theta^\alpha_\mu
d\widehat{\theta}^\alpha_\mu}{2\pi}\exp\Big[i\widehat{\theta}^\alpha_\mu\Big(\theta^\alpha_\mu-\frac{1}{\sqrt{N}}\sum_{j=1}^N
K^\alpha_j(\bxi_j^\mu)^2\Big)\Big]
\end{eqnarray}
to take the local fields out of the characteristic function and obtain
\begin{eqnarray}
\Odiss\prod_{\alpha=1}^n\prod_{\mu=1}^p\chi_{\xi^\mu_0 }(h ^\alpha_\mu,\theta ^\alpha_\mu;\kappa)\Cdiss=\int \Big[ \prod_{\alpha=1}^n\prod_{\mu=1}^p\frac{d \theta^\alpha_\mu d\widehat{\theta}^\alpha_\mu}{2\pi}\frac{d h^\alpha_\mu d\widehat{h}^\alpha_\mu}{2\pi}\Big] \exp\Big[i\sum_{\alpha=1}^n\sum_{\mu=1}^p\big(\widehat{h}^\alpha_\mu h^\alpha_\mu+\widehat{\theta}^\alpha_\mu\theta^\alpha_\mu\big)\Big]\nonumber\\
\Odiss\prod_{\alpha=1}^n\prod_{\mu=1}^p\exp\Big[-\frac{i\widehat{h}^\alpha_\mu}{\sqrt{N}}\sum_{j=1}^N J_j^\alpha\xi^\mu_j-\frac{i\widehat{\theta}^\alpha_\mu}{\sqrt{N}}\sum_{j=1}^N  K^\alpha_j(\xi^\mu_j)^2\Big]\Cdiss_{\xi_i^\mu}\Odiss\prod_{\alpha=1}^n\prod_{\mu=1}^p\chi_{\xi^\mu_0
}(h ^\alpha_\mu,\theta ^\alpha_\mu;\kappa)\Cdiss_{\xi_o^\mu} \, .
\end{eqnarray}
Introducing the order parameters (\ref{orderpar})and their conjugate variables, and using the identities
\begin{eqnarray}
1=\int_{-\infty}^\infty \prod_{\alpha<\beta}\frac{d q_{\alpha\beta}d\widehat{q}_{\alpha\beta}}{2\pi i/N}\exp\Big[\widehat{q}_{\alpha\beta}\Big(Nq_{\alpha\beta}- {\bf J}^{\alpha}\cdot{\bf J}^{\beta}\Big)\Big]\\
1=\int_{-\infty}^\infty \prod_{\alpha<\beta}\frac{d r_{\alpha\beta}d\widehat{r}_{\alpha\beta}}{2\pi i/N}\exp\Big[\widehat{r}_{\alpha\beta}\Big(Nr_{\alpha\beta}- {\bf K}^{\alpha}\cdot {\bf K}^{\beta}\Big)\Big]\\
1=\int_{-\infty}^\infty\prod_{\alpha=1}^n\frac{d
L^{\alpha}d\widehat{L}^{\alpha}}{2\pi /\sqrt{N}}
\exp\Big[i\widehat{L}^{\alpha}\Big(\sqrt{N}L^{\alpha}-\sum_{j=1}^N
K^{\alpha}_{j}\Big)\Big]
\end{eqnarray}
allows us to express the replicated fractional volume as an integral over them, viz.
\begin{equation}
\odis V^n\cdis\propto\int_{-\infty}^\infty\Big[\prod_{\alpha=1}^n
\frac{d L^{\alpha}d\widehat{L}^{\alpha}}{2\pi/\sqrt{N}}\Big]\Big[\prod_{\alpha=1}^n
\frac{d\widehat{E}^\alpha}{4\pi i}\frac{d\widehat{F}^\alpha}{4\pi i}\Big]\Big[\prod_{\alpha<\beta}\frac{d q_{\alpha\beta}d\widehat{q}_{\alpha\beta}}{2\pi i/N}\frac{d
r_{\alpha\beta}d\widehat{r}_{\alpha\beta}}{2\pi i/N}\Big]\exp\Big[N\Phi\Big]
 \label{volume2}
\end{equation}
with $\Phi$ given by
\begin{equation}
\Phi=\alpha G_1(q_{\alpha\beta},r_{\alpha\beta},L^\alpha)+G_2(\widehat{Q}_{\alpha\beta},\widehat{R}_{\alpha\beta},\widehat{L}^\alpha)+G_3(q_{\alpha\beta},r_{\alpha\beta},\widehat{Q}_{\alpha\beta},\widehat{ R}_{\alpha\beta})
\label{Phifunction}
\end{equation}
where
\begin{eqnarray}
G_1=&&\ln\int_{-\infty}^\infty \Big[\prod_{\alpha=1}^n\frac{d \theta^\alpha d\widehat{\theta}^\alpha}{2\pi}\frac{d h^\alpha d\widehat{h}^\alpha}{2\pi}\Big]\exp\Big[i\sum_{\alpha=1}^n(\widehat{h}^\alpha h^\alpha+\widehat{\theta}^\alpha\theta^\alpha) -ia\sum_{\alpha=1}^n\widehat{\theta}^\alpha L^\alpha -\frac{a}{2} \sum_{\alpha,\beta=1}^n\widehat{h}^\alpha\widehat{h}^\beta q_{\alpha\beta}\nonumber\\
&&-\frac{a(1-a)}{2} \sum_{\alpha,\beta=1}^n\widehat{\theta}^\alpha\widehat{\theta}^\beta r_{\alpha\beta}\Big]\Odiss\prod_{\alpha=1}^n\chi_{\xi }(h ^\alpha,\theta ^\alpha;\kappa)\Cdiss_{\xi_o}\\
G_2=&&\ln\int_{-\infty}^\infty \Big[\prod_{\alpha=1}^n d J^\alpha d K^\alpha\Big]\exp\Big[-\frac{1}{2}\sum_{\alpha,\beta=1}^n\Big(\widehat{Q}_{\alpha\beta}J^{\alpha}J^{\beta}+\widehat{R}_{\alpha\beta} K^{\alpha}K^{\beta}\Big)-i\sum_{\alpha=1}^n\widehat{L}^{\alpha} K^{\alpha}\Big]\\
G_3=&&\frac{1}{2}\sum_{\alpha,\beta=1}^n\Big(\widehat{Q}_{\alpha\beta}Q_{\alpha\beta}+\widehat{R}_{\alpha\beta}R_{\alpha\beta}\Big)
\end{eqnarray}
\end{widetext}
and 
\begin{eqnarray}
\widehat{Q}_{\alpha\beta}=\widehat{E}^\alpha\delta_{\alpha\beta}+\widehat{q}_{\alpha\beta}(1-\delta_{\alpha\beta})\\
\widehat{R}_{\alpha\beta}=\widehat{F}^\alpha\delta_{\alpha\beta}+\widehat{r}_{\alpha\beta}(1-\delta_{\alpha\beta})\\
Q_{\alpha\beta}=\delta_{\alpha\beta}+q_{\alpha\beta}(1-\delta_{\alpha\beta})\\
R_{\alpha\beta}=\delta_{\alpha\beta}+r_{\alpha\beta}(1-\delta_{\alpha\beta}) \, .
\end{eqnarray}
We remark that the $\delta$-function representation of the local fields has allowed us to perform the calculations until this point without using an explicit form for the characteristic
function $\chi_{\xi }(h ^\alpha,\theta ^\alpha;\kappa)$. 
Using the RS ansatz $\Phi$ can be simplified further and the saddle-point equations for ${\hat Q},Q,{\hat R},R$ become algebraic so that they can be solved explicitly, leading to the result (\ref{capacity_RS2})-(\ref{gRS2}).
\section{$q \to 1$ limit}
\begin{figure}
\psfrag{c}{$y$} 
\psfrag{d}{$x$} 
\psfrag{b}{$\Omega_{\xi=1}$} 
\psfrag{a}{${\cal R}_1$} 
\psfrag{e}{${\cal R}_2$} 
\psfrag{f}{${\cal R}_3$}
\psfrag{punto1}{$(h_0,\theta_0)$} 
\psfrag{punto2}{$(h_0,\theta_0)$} 
\psfrag{punto3}{$(h_0,\theta_0)$} 
\includegraphics[width=0.38\textwidth,height=0.28\textheight]{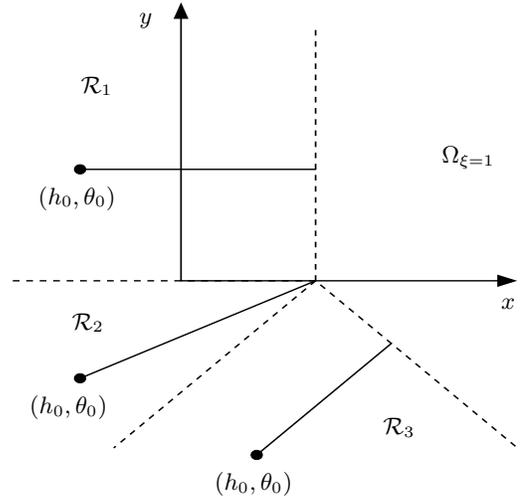}
\caption{Schematic representation of the subregions and minimal distances for ${\cal C}(\Omega_\xi=1)$ }
\label{figure7}
\end{figure}

In order to compute the asymptotic expansion of (\ref{gRS2}) we
proceed as follows. We split the integral over $(h_0,\theta_0)$ into
two parts, i.e., $\Omega_\xi$ determined by the Heaviside function in $\chi_\xi$, and its complement ${\cal C}(\Omega_\xi)$.  The first integral gives zero contribution in the limit $q\to1 $, while the second one gives a contribution of order $(1-q)^{-1}$.
Indeed, the integration over $(h,\theta)$ parametrized by $q$ is nothing
but an exponential Dirac-delta representation. Whenever the peak of this
delta representation lies in the region $\Omega_\xi$, which means that
$(h_0,\theta_0)\in\Omega_\xi$  the integral results in a finite
contribution. The contributions of order $(1-q)^{-1}$ arises from the
points  $(h_0,\theta_0)\in{\cal C}(\Omega_\xi)$.
Therefore, we can write
\begin{equation}
g^{RS}_1(q,\gamma,L)=\Odiss\int_{{\cal C}(\Omega_\xi)}
   {\cal D}(h_0) {\cal D}(\sqrt{\gamma} \theta_0-l)
      \ln[1]_\xi(h_0,\theta_0)\Cdiss_\xi
\end{equation}
where we have introduced the shorthand notation
\begin{eqnarray}
[1]_\xi(h_0,\theta_0)&&=\int_{\Omega_\xi}\frac{dh}{\sqrt{2\pi(1-q)}} \frac{d\theta}{\sqrt{2\pi(1-q)}}\nonumber\\
   && \times \exp\Big[-\frac{(h-h_0)^2+(\theta-\theta_0)^2}{2(1-q)}\Big]          \, .
   \label{func1}
\end{eqnarray} 
Next, for a given $(h_0,\theta_0)\in{\cal C}(\Omega_\xi)$ the
main contribution arising from the function $[1]_\xi(h_0,\theta_0)$
is obtained for those points $(h,\theta)\in\Omega_\xi$ which minimize
the distance $(h-h_0)^2+(\theta-\theta_0)^2$. To calculate this minimal
distance, we split up ${\cal C}(\Omega_\xi)$ into three subregions according to fig. \ref{figure7} in the case of $\xi=1$
\begin{eqnarray}
&& \hspace*{-0.8cm}{\cal R}_1=\left\{\begin{array}{l}
h_0<\frac{\kappa}{\sqrt{a}}\\
\theta_0>0
\end{array}\right.\\
&&\hspace*{-0.8cm}{\cal R}_2=\left\{\begin{array}{l}
\Big(h_{0}-\frac{\kappa}{\sqrt{a}}\Big)\sqrt{\gamma  (1-a)}<\theta_{0}<0\\ 
h_{0}<\frac{\kappa}{\sqrt{a}}
\end{array}\right.\\
&&\hspace*{-0.8cm}{\cal R}_3=\left\{\begin{array}{l}
\theta_{0}<0\\ 
\frac{\theta_{0}}{\sqrt{\gamma  (1-a)}}+\frac{\kappa}{\sqrt{a}}<h_{0}<\frac{\kappa}{\sqrt{a}}-\theta_{0}\sqrt{\gamma  (1-a)}  \, . \\
\end{array}\right.
\end{eqnarray}

\begin{figure}
\psfrag{theta}{$y$} 
\psfrag{h}{$x$} 
\psfrag{omega}{$\Omega_{\xi=0}$} 
\psfrag{r1}{${\cal R}'_1$} 
\psfrag{r2}{${\cal R}'_2$} 
\psfrag{r3}{${\cal R}'_3$}
\psfrag{punto1}{$(h_0,\theta_0)$} 
\psfrag{punto2}{$(h_0,\theta_0)$} 
\psfrag{punto3}{$(h_0,\theta_0)$} 
\includegraphics[width=0.38\textwidth,height=0.28\textheight]{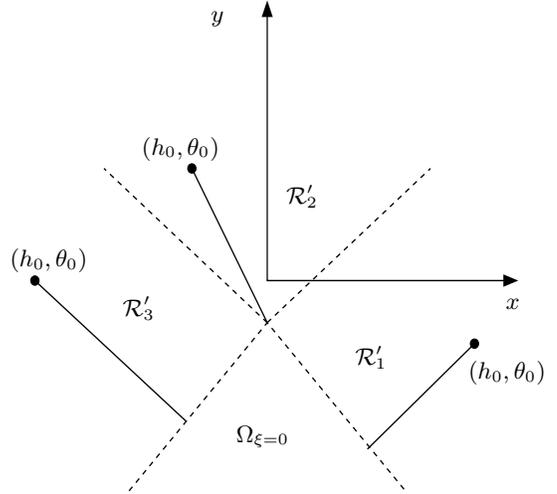}
\caption{As figure \ref{figure7} for ${\cal C}(\Omega_\xi=0)$ }
\label{figure8}
\end{figure}

Computing the minimal distances for such subregions is straightforward and leads to
\begin{eqnarray}
&&\hspace*{-0.5cm}d^{{\cal R}_1}_{min}=\Big(\frac{\kappa}{\sqrt{a}}-h_0\Big)^2\\
&&\hspace*{-0.5cm}d^{{\cal R}_2}_{min}=\Big(\frac{\kappa}{\sqrt{a}}-h_0\Big)^2 +\theta_0^2\\
&&\hspace*{-0.5cm}d^{{\cal R}_3}_{min}=\frac{1}{1+\gamma(1-a)}\Big(\frac{\kappa}{\sqrt{a}}-\theta_0\sqrt{\gamma(1-a)}-h_0\Big)^2 \, .
\end{eqnarray}
By redefining $h_0-\kappa/\sqrt{a}\rightarrow h_0$ and $\gamma'=\sqrt{\gamma(1-a)}$ we recover the expressions (\ref{Mindis1})-(\ref{Mindis3}).\\
We proceed analogously for the region ${\cal C}(\Omega_{\xi=0})$.
We split this region into three subregions as shown in fig. \ref{figure8}
\begin{widetext}
\begin{eqnarray}
{\cal R}'_1&&=\left\{\begin{array}{l}
h_{0}>0\\
-\frac{1}{\sqrt{\gamma (1-a)}}\Big(h_{0}+\frac{\kappa}{\sqrt{a}}\Big)<\theta_{0}<\sqrt{\gamma (1-a)}h_{0}-\frac{\kappa}{\sqrt{\gamma a(1-a)}}
\end{array}\right.\\
{\cal R}'_2&&=\left\{\begin{array}{l}
-\frac{1}{\sqrt{\gamma (1-a)}}\Big(\theta_{0}+\frac{\kappa}{\sqrt{\gamma a(1-a)}}\Big)<h_{0}<\frac{1}{\sqrt{\gamma (1-a)}}\Big(\theta_{0}+\frac{\kappa}{\sqrt{\gamma a(1-a)}}\Big)\\ 
-\frac{\kappa}{\sqrt{\gamma a(1-a)}}<\theta_{0}<\infty 
\end{array}\right.\\
{\cal R}'_1&&=\left\{\begin{array}{l}
h_{0}<0\\
-\frac{1}{\sqrt{\gamma (1-a)}}\Big(-h_{0}+\frac{\kappa}{\sqrt{a}}\Big)<\theta_{0}<-\sqrt{\gamma (1-a)}h_{0}-\frac{\kappa}{\sqrt{\gamma a(1-a)}} \, .
\end{array}\right.
\end{eqnarray}
\end{widetext}
The minimal distances are given by 
\begin{eqnarray}
&&  d^{{\cal R}'_1}_{min}=\frac{\big(\sqrt{\gamma a(1-a)}\theta_{0}+\kappa+\sqrt{a}h_{0}\big)^2}{a[1+\gamma(1-a)]}\\
&&  d^{{\cal R}'_2}_{min}=h^2_{0}+\Big(\frac{\kappa}{\sqrt{\gamma a(1-a)}}+\theta_{0}\Big)^2\\
&&  d^{{\cal R}'_3}_{min}=\frac{\big(\sqrt{\gamma a(1-a)}\theta_{0}+\kappa-\sqrt{a}h_{0}\big)^2}{a[1+\gamma(1-a)]}
\end{eqnarray}
and redefining $\kappa/\sqrt{\gamma a(1-a)}+\theta_{0}\rightarrow \theta_0$ we find (\ref{Mindis1prim})-(\ref{Mindis3prim}).

\section{RS stability}
Starting from the stability matrix formed by the second derivatives of $\Phi$ (recall eq.(\ref{Phifunction})) with respect to the order parameters and the conjugated variables, we find that only transverse fluctuations are relevant.

These transverse fluctuations are characterized by four eigenvalues with degeneracy $n(n-3)/2$, given by the roots of the fourth degree characteristic polynomial $P(\lambda)$
\begin{widetext}
\begin{equation}
P(\lambda)=\begin{vmatrix}
\Delta_q-\lambda&\Delta_c&1&0\\
\Delta_c&\Delta_r-\lambda&0&1\\
1&0&\Delta_{\widehat{q}}-\lambda&0\\
0&1&0&\Delta_{\widehat{r}}-\lambda
\end{vmatrix}=[(\Delta_{q}-\lambda)(\Delta_{\widehat{q}}-\lambda)-1][(\Delta_{r}-\lambda)(\Delta_{\widehat{r}}-\lambda)-1]-\Delta_{c}^2(\Delta_{\widehat{q}}-\lambda)(\Delta_{\widehat{r}}-\lambda)
\end{equation}
\end{widetext}
with the coefficients $\Delta$ given by
\begin{widetext}
\begin{eqnarray}
&&\Delta_q=
  \frac{\alpha}{q^2} \int{\cal D}(h_0){\cal D}(\sqrt{\gamma}\theta_0-t)
         \Odiss\Big\{\frac{\partial^2}{\partial                                  h_0^2}\ln[1]_\xi(h_0,\theta_0)\Big\}^2\Cdiss_{\xi_o}\\
&&\Delta_r=
  \frac{\alpha}{r^2}\int{\cal D}(h_0){\cal D}(\sqrt{\gamma}\theta_0-t)
\Odiss\Big\{\frac{1}{\gamma}\frac{\partial^2}{\partial \theta_0^2}\ln[1]_\xi(h_0,\theta_0)\Big\}^2\Cdiss_{\xi_o}\\
&&\Delta_c=
 \frac{\alpha}{qr}\int{\cal D}(h_0){\cal D}(\sqrt{\gamma}\theta_0-t)
\Odiss\Big\{\frac{1}{\sqrt{\gamma}}\frac{\partial^2}{\partial h_0\partial \theta_0}\ln[1]_\xi (h_0,\theta_0)\Big\}^2\Cdiss_{\xi_o}\\
&&\Delta_{\widehat{q}}=(1-q)^2\\
&&\Delta_{\widehat{r}}=(1-r)^2=\gamma^2(1-q)^2
\label{deltas}
\end{eqnarray}
\end{widetext}
where we recall that $(1-r)=\gamma(1-q)$
and the function $[1]_\xi (h_0,\theta_0)$ is defined in (\ref{func1}).

Next, the limit $q\to1$ has to be taken. 
Using the asymptotic expansion of $[1]_\xi(h_0,\theta_0)$ discussed in
appendix B we can compute the asymptotic behavior of
 the coefficients $\Delta_q$, $\Delta_r$ and $\Delta_c$. After a lot of algebra we finally arrive at the expressions (\ref{delq})-(\ref{delr})
with the integration regions and minimal distances given by (\ref{Region1})-(\ref{Mindis3prim}). In this limit, it turns out that an analytical expression can be found for the eigenvalues.
First, we notice that the determinant of the matrix remains
finite in the limit. Since the determinant is the product of the eigenvalues, it follows that this  product is finite.
Two possibilities arise, either all eigenvalues are finite, or
two of them tend to zero and two to infinity with the same ratio.
It is not hard to prove that the first choice is incorrect. Hence, two of the eigenvalues have to behave asympotically as
$(1-q)^{\pm n}$. One can check that only $n=2$ is possible.
This allows us to split $P(\lambda)$ into two polynomials which give the solutions around zero and around infinity. These polynomials read
\begin{eqnarray}
P_0(\lambda)=[\Delta_{q}(\Delta_{\widehat{q}}-\lambda)-1][\Delta_{r}(\Delta_{\widehat{r}}-\lambda)-1] \nonumber\\
-\Delta_{c}^2(\Delta_{\widehat{q}}-\lambda)(\Delta_{\widehat{r}}-\lambda) \nonumber\\
P_\infty(\lambda)=(\Delta_{q}-\lambda)(\Delta_{r}-\lambda)-\Delta_{c}^2 \, .
\end{eqnarray}
From these two polynomials the four eigenvalues (\ref{lplus})-(\ref{tmin}) can be found. We remark that in the limit $a \to 1$ we find back the stability criteria for the original Gardner capacity problem.


\end{document}